\documentclass[aip,jcp,11pt]{revtex4-1}

\draft
\usepackage{amsmath}
\usepackage{amssymb}
\usepackage{graphicx}
\usepackage{esint}
\PassOptionsToPackage{version=3}{mhchem}
\usepackage{mhchem}
\usepackage[unicode=true,pdfusetitle,
 bookmarks=true,bookmarksnumbered=false,bookmarksopen=false,
 breaklinks=false,pdfborder={0 0 1},backref=section,colorlinks=false]
 {hyperref}

\makeatletter

\usepackage{amsthm}% Useful AMS stuff
\usepackage{tikz}\usetikzlibrary{arrows}
\usepackage{caption}
\usepackage{subcaption}
\usepackage{color}

\usepackage{multirow}\usepackage{bigdelim}

\usepackage[section]{placeins}

\newcommand{\bra}[1]{\langle #1 \mid}
\newcommand{\ket}[1]{\mid #1 \rangle}
\newcommand{\bracket}[2]{\langle #1 \mid #2 \rangle}
\usepackage{bbold}\newcommand{\identity}{\mathbb{1}}

\usepackage{mathtools}

\renewcommand{\imath}{\text{\rm{i}}}

\usepackage{titlesec}\titleformat{\chapter}
{\normalfont\huge\bfseries}{\chaptertitlename\ \thechapter:\ \ }{0em}{} 
\titlespacing*{\chapter}{0pt}{0pt}{30pt}

\begin{document}

%%%%%%%%%%%%%%%%%%%%%%%%%%%%%%%%%%%%%%%%%%%%%%%%%%%%%%%%%%%%%%%%%%%%%
%%%%%%%%%%%%%%%%%%%%%%%%%%%%%%%%%%%%%%%%%%%%%%%%%%%%%%%%%%%%%%%%%%%%%
%% Meta-data block
%% ---------------
%% Each author should be given as a separate \author command.
%%
%% Corresponding authors should have an e-mail given after the author
%% name as an \email command. Phone and fax numbers can be given
%% using \phone and \fax, respectively; this information is optional.
%%
%% The affiliation of authors is given after the authors; each
%% \affiliation command applies to all preceding authors not already
%% assigned an affiliation.
%%
%% The affiliation takes an option argument for the short name.  This
%% will typically be something like "University of Somewhere".
%%
%% The \altaffiliation macro should be used for new address, etc.
%% On the other hand, \alsoaffiliation is used on a per author basis
%% when authors are associated with multiple institutions.
%%%%%%%%%%%%%%%%%%%%%%%%%%%%%%%%%%%%%%%%%%%%%%%%%%%%%%%%%%%%%%%%%%%%%
\author{Alexander Humeniuk}
\author{Roland Mitri\'{c}}
\affiliation{Institut f\"{u}r Physikalische und Theoretische Chemie, Julius-Maximilians Universit\"{a}t W\"{u}rzburg, Emil-Fischer-Stra\ss e 42, 97074 W\"{u}rzburg}
\email{roland.mitric@uni-wuerzburg.de}
%%%%%%%%%%%%%%%%%%%%%%%%%%%%%%%%%%%%%%%%%%%%%%%%%%%%%%%%%%%%%%%%%%%%%
%% The document title should be given as usual. Some journals require
%% a running title from the author: this should be supplied as an
%% optional argument to \title.
%%%%%%%%%%%%%%%%%%%%%%%%%%%%%%%%%%%%%%%%%%%%%%%%%%%%%%%%%%%%%%%%%%%%%
\title{Non-Adiabatic Dynamics around a Conical Intersection with Surface-Hopping Coupled Coherent States}

%%%%%%%%%%%%%%%%%%%%%%%%%%%%%%%%%%%%%%%%%%%%%%%%%%%%%%%%%%%%%%%%%%%%%
%% Some journals require a list of abbreviations or keywords to be
%% supplied. These should be set up here, and will be printed after
%% the title and author information, if needed.
%%%%%%%%%%%%%%%%%%%%%%%%%%%%%%%%%%%%%%%%%%%%%%%%%%%%%%%%%%%%%%%%%%%%%
%\abbreviations{}
%\keywords{}

%%%%%%%%%%%%%%%%%%%%%%%%%%%%%%%%%%%%%%%%%%%%%%%%%%%%%%%%%%%%%%%%%%%%%
%% The manuscript does not need to include \maketitle, which is
%% executed automatically.
%%%%%%%%%%%%%%%%%%%%%%%%%%%%%%%%%%%%%%%%%%%%%%%%%%%%%%%%%%%%%%%%%%%%%

\makeatother

%%%%%%%%%%%%%%%%%%%%%%%%%%%%%%%%%%%%%%%%%%%%%%%%%%%%%%%%%%%%%%%%%%%%%
%% The abstract environment will automatically gobble the contents
%% if an abstract is not used by the target journal.
%%%%%%%%%%%%%%%%%%%%%%%%%%%%%%%%%%%%%%%%%%%%%%%%%%%%%%%%%%%%%%%%%%%%%

\begin{abstract}
An extension of the CCS-method [\textit{Chem. Phys.} \textbf{2004,} \textsl{304,} 103-120] for simulating non-adiabatic dynamics with quantum effects of the nuclei is put forward.

The time-dependent Schr\"{o}dinger equation for the motion of the nuclei is solved
in a moving basis set. The basis set is guided by classical trajectories, which can hop stochastically between different electronic potential energy surfaces. The non-adiabatic transitions are modelled by a modified version of Tully's fewest switches algorithm. The trajectories consist of Gaussians in the phase space of the nuclei (coherent states) combined with amplitudes for an electronic wave function. The time-dependent matrix elements between different coherent states determine the amplitude of each trajectory in the total multistate wave function; the diagonal matrix elements determine the hopping probabilities and gradients. In this way, both intereference effects and non-adiabatic transitions can be described in a very compact fashion, leading to the exact solution if convergence with respect to the number of trajectories is achieved and the potential energy surfaces are known globally. 

The method is tested on a 2D model for a conical intersection [\textit{J. Chem. Phys.,} \textbf{1996,} \textsl{104,} 5517], where a nuclear wavepacket encircles the point of degeneracy between two potential energy surfaces and intereferes with itself. These intereference effects are absent in classical trajectory-based molecular dynamics but can be fully incorporated if trajectories are replaced by surface hopping coupled coherent states.
\end{abstract}
\maketitle
%%%%%%%%%%%%%%%%%%%%%%%%%%%%%%%%%%%%%%%%%%%%%%%%%%%%%%%%%%%%%%%%%%%%%
%% Start the main part of the manuscript here.
%%%%%%%%%%%%%%%%%%%%%%%%%%%%%%%%%%%%%%%%%%%%%%%%%%%%%%%%%%%%%%%%%%%%%

\section{Introduction}

In photochemistry quantum effects of the nuclei usually are only of
minor importance, while the electronic structure is decisive. That
is why classical molecular dynamics (in combination with surface hopping
to allow for electronic transitions)\cite{marx_md_on_excited_states_review}
has been quite successful in describing photochemical reactions. Nonetheless,
some exceptions to this exist where nuclear quantum effects are noticable
even at room temperature: The first is tunneling of light elements
such as hydrogen\cite{voth_H_tunneling}, and the second concerns
geometric phases that arise when potential energy surfaces (PES) become
degenerate at so-called conical intersections \cite{molecular_Akharonov_Bohm}
(molecular Akharonov-Bohm effect).

Conical intersections (CI) \cite{ci_book,diabolic_ci} are topological
features of the potential energy surfaces and thus remain equally
important at high as at low temperatures. They are the ``transition
states'' of photochemical reactions and interference
effects in the wake of a CI can determine the product ratio following
a radiationless internal conversion\cite{geometric_phase_phenol}.

If one is specifically interested in studying these nuclear effects,
classical molecular dynamics is not sufficient. Still one should not
abandon the concept of trajectories, for they have appealing advantages
over grid-based solutions of the Schrödinger equation: 
\begin{itemize}
\item Each trajectory and its hops between electronic states can be interpreted
as a photochemical reaction path. 
\item Trajectories automatically sample the interesting part of the nuclear
phase space and electronic state manifold. 
\end{itemize}
How can one include quantum-mechanical effects, while retaining a
trajectory-based description? The missing ingredients become evident
by comparison with Feynman's path integral formulation: The propagator
is obtained by summing over \textsl{all} paths weighted with a \textsl{phase}.
Therefore, 
\begin{itemize}
\item trajectories have to be allowed to explore more than the classically
allowed phase space, and 
\item they have to be equipped with a phase so that they can interfere. 
\end{itemize}
The coupled coherent states (CCS) method\cite{CCS_method} developed
by Shalashilin and Child fulfils these requirements. Trajectories
are replaced by coherent states similar to the frozen Gaussians \cite{frozen_gaussians}
introduced by Heller. They move classically on potential energy surfaces,
which, due to the finite width of the coherent states, are smoothed
out, so that the trajectories can access a larger phase-space volume.
The evolution of the phases attributed to the trajectories are computed
from the matrix elements of the nuclear hamiltonian between the coherent
state wavepackets. The phase of one trajectory depends on all the
others, so that the trajectories have to be propagated in parallel.
In this sense, quantum effects can be thought of as arising from the
interaction of the trajectories.

Non-adiabatic dynamics using coupled coherent states have been performed
before with the Ehrenfest method \cite{CCS_ehrenfest}. Here, a different
procedure is proposed, in which the trajectories do not move on the
average potential energy surface, but can hop stochastically between
different surfaces according to Tully's procedure for assigning the
hopping probabilities \cite{Tully_hopping_probabilities}. This approach
bears some resemblance to the method of surface hopping Gaussians
(SHG) by Horenko et.al. \cite{surface_hopping_gaussians}, however
being derived from the CCS-method, the working equations are different,
in particular the trajectories move on potentials that differ from
the classical ones due to the finite width of the coherent states.

The CCS method belongs to a wider class of methods, which solve the
Schrödinger equation in a time-dependent basis set:

% MCTDHThe oldest one of them is the multi-configurational time-dependent
Hartree (MCTDH) method\cite{mctdh_book,mctdh_zundel_cation}. Both
the time-evolution of the basis vectors and the coefficients is determined
from a variational principle. In MCTDH, the wavefunction is represented
by products of 1D functions, which can move along the axes so as to
track the wavepacket optimally.

% AIMSIn the ab-initio multiple spawning methods (AIMS)\cite{ab_initio_multiple_spawning},
the moving basis also consists of Gaussians. The basis is expanded
dynamically during non-adiabatic events, so that a wavepacket travelling
through a region of strong non-adiabatic coupling can split into several
Gaussians moving on different surfaces\cite{aims_ci}. Unlike in CCS,
the trajectories move on the classical potential energy surface, which
complicates the discription of tunneling, unless a special procedure
is included for spawning new trajectories on the other side of the
barrier\cite{aims_tunneling}. Recently, also a combination of AIMS
and CCS has been published \cite{ab_initio_multiple_cloning}.

% GridsThese methods have to be contrasted with techniques where
the wavefunction is represented on a set of regularly arranged mesh
points. The computational cost of wavepacket dynamics on a grid scales
steeply with the number of dimensions. In order to reduce the number
of dimensions, special coordinate systems\cite{Na3F_qmdynamics,wavepacket_Jacobi}
can be chosen, but the accompanying coordinate transformation leads
to a complicated form of kinetic operator, which is special to each
coordinate system. Essentially each molecular system requires a special
treatment. As opposed to this, trajectory-based wavepackets dynamics
can be performed in cartesian coordinates \cite{CCS_cartesian}, so
that the kinetic operator retains its simple form.

Trajectory-guided basis sets results in favourable scaling but slow
convergence, although methods have been developed to improve the sampling
of phase space\cite{ccs_sampling}. If the trajectories spread too
quickly in phase space coupling between the trajectories is lost.
From an unconverged CCS simulation with surface hopping trajectories,
useful information can still be extracted. This is less the case for
Ehrenfest dynamics, where an individual trajectory has no intuitive
meaning.

% PESMolecular dynamics simulations require accurate potential energy
surfaces, and the way these are obtained lead to some restrictions.
Ab-initio quantum chemistry methods solve for the electronic structure
at a fixed nuclear geometry. \textsl{Direct dynamics} only requires
energies, gradients and non-adiabatic couplings, which are calculated
along each trajectory ``on the fly''. SHG\cite{surface_hopping_gaussians},
AIMS\cite{aims_molpro} and MCTDH\cite{mctdh_nonadiabatic,burghardt_vMCG}
have been adapted to be compatible with quantum chemistry methods
by approximating the matrix elements between different trajectory
wavepackets only by local quantities available at each trajectory
position. This makes them suitable for large, complicated systems,
but the price to be paid is that the description becomes only semiclassical.
Even if the trajectories are coupled, the approximate phases do not
result in the correct interference pattern.

Currently, it seems that exact quantum dynamics can only be achieved
if the potential energy surfaces are known globally. Fitting entire
surfaces is only feasible for very small molecules\cite{truhlar_fit_NH3}.
Parts of the surface, e.g. the region around a conical intersection,
can be fitted to ab-initio calculations in the form of a vibronic
coupling hamiltonian\cite{ci_book}. Another approach consists in
using model potentials. In principle, complex diabatic potentials
can be constructed from basic building blocks for which the matrix
elements can be computed analytically in the spirit of force fields.
This will be the path followed here.

\textbf{Outline of the article:} First the modified CCS algorithms
is described, that allows trajectories to switch between potential
energy surfaces if a change of the electronic wave function is detected.
The equations of motion for the moving basis set and the phases are
derived. Finally the scattering of a wave packet off the 2D model
of a conical intersection\cite{ci_model} is explored using the CCS
method with surface hopping trajectories. Comparison with the numerically
exact solution shows that the interference effects can be fully reproduced.

% METHOD DESCRIPTION

\section{Method Description}

\subsection{Schrödinger's equation in a moving basis set}

The goal is to solve the time-dependent Schrödinger equation for a
diabatic Hamiltonian with $N_{\text{dim}}$ nuclear degrees of freedom
and $N_{\text{st}}$ electronic states, 
\begin{equation}
\imath\hbar\frac{d}{dt}\Psi_{A}\left(x_{1},\ldots,x_{N_{dim}}\right)=\sum_{B=1}^{N_{\text{st}}}H_{AB}\left(\hat{q}_{1},\ldots,\hat{q}_{N_{\text{dim}}};\hat{p}_{1},\ldots,\hat{p}_{N_{\text{dim}}}\right)\Psi_{B}\left(x_{1},\ldots,x_{N_{\text{dim}}}\right),
\end{equation}
in a moving basis set. In the following $A,B,I$ and $J$ will be
used to label electronic states, $i,j$ and $k$ will label basis
vectors and $d$ enumerates the nuclear dimensions.

Wavepacket dynamics can be tracked efficiently if the wave function
is expanded into a set of moving basis functions \cite{CCS_method,CCS_ehrenfest,CCS_cartesian,footnote_moving_basis}.
A convenient choice of basis functions for the nuclear degrees of
freedom are \textbf{coherent states} $\ket{\mathbf{z}}$, whose position
representation is given by\cite{CCS_method} 
\begin{equation}
\bracket{\mathbf{x}}{\mathbf{z}}=\left(\frac{\gamma}{\pi}\right)^{N_{\text{dim}}/4}\exp\left(\sum_{d=1}^{N_{\text{dim}}}\left[-\frac{\gamma}{2}(x_{d}-q_{d})^{2}+\frac{\imath}{\hbar}p_{d}(x_{d}-q_{d})+\frac{\imath}{\hbar}p_{d}q_{d}\right]\right)
\end{equation}
where $\gamma$ is an adjustable parameter that controls the spatial
width of the coherent state. A coherent state is labelled by a complex
$D_{\text{dim}}$-dimensional vector $\mathbf{z}=\sqrt{\frac{\gamma}{2}}\mathbf{q}+\frac{\imath}{\hbar}\sqrt{\frac{1}{2\gamma}}\mathbf{p}$,
where $\mathbf{q}$ and $\mathbf{p}$ are the coordinates of its maximum
amplitude in phase space. Coherent states are right eigen vectors
of the scaled annihilation operator $\hat{\mathbf{a}}$ and left eigen
vectors of the scaled creation operator $\hat{\mathbf{a}}^{\dagger}$\cite{CCS_method}:
\begin{eqnarray}
\hat{\mathbf{a}} & = & \sqrt{\frac{\gamma}{2}}\hat{\mathbf{q}}+\frac{\imath}{\hbar}\sqrt{\frac{1}{2\gamma}}\hat{\mathbf{p}}\\
\hat{\mathbf{a}}^{\dagger} & = & \sqrt{\frac{\gamma}{2}}\hat{\mathbf{q}}-\frac{\imath}{\hbar}\sqrt{\frac{1}{2\gamma}}\hat{\mathbf{p}}\\
\hat{\mathbf{a}}\ket{\mathbf{z}} & = & z\ket{\mathbf{z}}\\
\bra{\mathbf{z}}\hat{\mathbf{a}}^{\dagger} & = & z^{*}\bra{\mathbf{z}}\\
\end{eqnarray}
\textbf{Matrix elements} of an operator $\hat{O}$ between coherent
states are particularly simple if the canonical position and momentum
operators $\hat{\mathbf{q}}=\sqrt{\frac{1}{2\gamma}}\left(\hat{\mathbf{a}}+\hat{\mathbf{a}}^{\dagger}\right)$
and $\hat{\mathbf{p}}=\sqrt{2\gamma}\frac{\hbar}{\imath}\left(\hat{\mathbf{a}}-\hat{\mathbf{a}}^{\dagger}\right)$
are expressed in terms of the creation and annihilation operators
and if the resulting products are brought into normal ordering (creation
operators preceed annihilation operators). The reordering is accomplished
by applying the commutation relation $\hat{\mathbf{a}}\hat{\mathbf{a}}^{\dagger}=\hat{\mathbf{a}}^{\dagger}\hat{\mathbf{a}}+\mathbb{1}$
repeatedly. 
\begin{eqnarray}
O(\hat{\mathbf{q}},\hat{\mathbf{p}}) & = & O_{\text{ord}}(\hat{\mathbf{a}}^{\dagger},\hat{\mathbf{a}})\\
\bra{\mathbf{z}_{1}}\hat{O}\ket{\mathbf{z}_{2}} & = & \bra{\mathbf{z}_{1}}\hat{O}_{\text{ord}}\ket{\mathbf{z}_{2}}=\bracket{\mathbf{z}_{1}}{\mathbf{z}_{2}}O_{\text{ord}}(\mathbf{z}_{1}^{*},\mathbf{z}_{2})\label{eqn:reordered_operator}
\end{eqnarray}
In practice, the reordered form of a potential $V(\mathbf{x})$ is
not obtained by algebraic reordering, but by solving the multidimensional
integral 
\begin{equation}
V_{\text{ord}}(\mathbf{z}_{1}^{*},\mathbf{z}_{2})=\int\bracket{\mathbf{z}_{1}}{\mathbf{x}}V(\mathbf{x})\bracket{\mathbf{x}}{\mathbf{z}_{2}}d^{N_{\text{dim}}}\mathbf{x}
\end{equation}
analytically, which is possible for a sufficiently large set of functions,
from which interesting model potentials can be constructed.

Coherent states are not orthogonal and form an overcomplete basis
of the Hilbert space\cite{klauder_book}: 
\begin{equation}
\bracket{\mathbf{z}_{1}}{\mathbf{z}_{2}}=\exp\left(\mathbf{z}_{1}^{*}\cdot\mathbf{z}_{2}-\frac{\vert\mathbf{z}_{1}\vert^{2}}{2}-\frac{\vert\mathbf{z}_{2}\vert^{2}}{2}\right)\label{eqn:zz_overlap}
\end{equation}
The identity operator is\cite{klauder_book}: 
\begin{equation}
\hat{Id}=\frac{1}{\pi}\int d^{2}\mathbf{z}\ket{\mathbf{z}}\bra{\mathbf{z}}
\end{equation}

In order to describe non-adiabatic dynamics, the basis vectors have
to span multiple states. A basis function thus consists of a nuclear
part, which is the same for all electronic states, and an electronic
part, which is represented by a $N_{\text{st}}$-dimensional complex
vector $\mathbf{a}$: 
\begin{equation}
\ket{\mathbf{z}_{i},\mathbf{a}_{i}}=\underbrace{\ket{\mathbf{z}_{i}}}_{\text{nuclear}}\otimes\underbrace{\sum_{A}a_{i}^{A}\ket{\chi_{A}(\mathbf{z}_{i})}}_{\text{electronic part}}
\end{equation}
Assuming that the electronic states $\ket{\chi_{A}}$ are diabatic
states, which do not change on the length scale where different coherent
states overlap, the overlap matrix between two coherent states with
electronic amplitudes can be calculated as: 
\begin{equation}
\Omega_{i,j}=\bracket{\mathbf{z}_{i},\mathbf{a}_{i}}{\mathbf{z}_{j},\mathbf{a}_{j}}=\bracket{\mathbf{z}_{i}}{\mathbf{z}_{j}}\sum_{A}\sum_{B}a_{i}^{A*}a_{j}^{B}\underbrace{\bracket{\chi_{A}(\mathbf{z}_{i})}{\chi_{B}(\mathbf{z}_{j})}}_{=\delta_{AB}}=\bracket{\mathbf{z}_{i}}{\mathbf{z}_{j}}\bracket{\mathbf{a}_{i}}{\mathbf{a}_{j}}\label{eqn:overlap_Omega_ij}
\end{equation}

If only a limited number of basis functions is used to describe the
Hilbert space in a region of interest, the discrete representation
of the identity has to be used \cite{CCS_method}: 
\begin{equation}
\identity=\sum_{i,j}\ket{\mathbf{z}_{i},\mathbf{a}_{i}}\left(\Omega^{-1}\right)_{ij}\bra{\mathbf{z}_{j},\mathbf{a}_{j}}\label{eqn:discrete_identity}
\end{equation}

By making the parameters of the basis functions time dependent, $\mathbf{z}_{i}\rightarrow\mathbf{z}_{i}(t),\mathbf{a}_{i}\rightarrow\mathbf{a}_{i}(t)$,
we obtain a moving basis set. The positions and momenta of the basis
functions will follow classical equations of motions on a reordered
potential, while the electronic coefficients $\mathbf{a}_{i}(t)$
determine the tendency of trajectories to hop to different surfaces.
While the dynamics of the basis functions is similar to Tully's surface
hopping, the coefficients of the wavefunction relative to the moving
basis and their coupling captures all quantum effects.

In what follows the differential equations governing the time-evolution
of the coefficients will be derived. The presentation
of the material follows reference \cite{CCS_method}, where the analogous
expressions for the single potential can be found.

The multistate wave function $\ket{\Psi}$ evolves according to Schrödinger's
equation: 
\begin{equation}
\imath\hbar\frac{\partial}{\partial t}\ket{\Psi}=\hat{H}\ket{\Psi}
\end{equation}

The hamiltonian $\hat{H}=\sum_{A,B}\ket{\chi_{A}}H_{AB}(\hat{\mathbf{q}},\hat{\mathbf{p}})\bra{\chi_{B}}$
can be reordered: 
\begin{equation}
H_{AB}(\hat{\mathbf{q}},\hat{\mathbf{p}})\rightarrow H_{AB}^{\text{ord}}(\hat{\mathbf{a}}^{\dagger},\hat{\mathbf{a}})
\end{equation}

First, the time-dependence of the projection of $\ket{\Psi}$ onto
the basis vector $i$ is considered. Since the basis vectors themselves
depend on time, the chain rules gives three terms (a dot is used to
denote a time derivative): 
\begin{equation}
\frac{d}{dt}\bracket{\mathbf{z}_{i},\mathbf{a}_{i}}{\Psi}=\bracket{\dot{\mathbf{z}}_{i},\mathbf{a}_{i}}{\Psi}+\bracket{\mathbf{z}_{i},\dot{\mathbf{a}}_{i}}{\Psi}+\bracket{\mathbf{z}_{i},\mathbf{a}_{i}}{\dot{\Psi}}
\end{equation}
Inserting the discrete identity, eqn. \ref{eqn:discrete_identity},
and the Schrödinger equation to replace $\ket{\dot{\Psi}}$ yields:
\begin{equation}
\begin{split}\frac{d}{dt}\bracket{\mathbf{z}_{i},\mathbf{a}_{i}}{\Psi}=\sum_{j,k}\Big\{ & \bracket{\dot{\mathbf{z}}_{i},\mathbf{a}_{i}}{z_{j},\mathbf{a}_{j}}+\bracket{\mathbf{z}_{i},\dot{\mathbf{a}}_{i}}{\mathbf{z}_{j},\mathbf{a}_{j}}\\
- & \frac{\imath}{\hbar}\bra{\mathbf{z}_{i},\mathbf{a}_{i}}\hat{H}\ket{\mathbf{z}_{j},\mathbf{a}_{j}}\Big\}\left(\Omega^{-1}\right)_{j,k}\bracket{\mathbf{z}_{k},\mathbf{a}_{k}}{\Psi}
\end{split}
\label{eqn:derivatives}
\end{equation}

After differentiating the overlap in eqn. \ref{eqn:zz_overlap} with
respect to the time-dependence of $\mathbf{z}_{1}$ and using relation
\ref{eqn:reordered_operator}, eqn. \ref{eqn:derivatives} becomes:
\begin{equation}
\begin{split}\frac{d}{dt}\bracket{\mathbf{z}_{i},\mathbf{a}_{i}}{\Psi}=\sum_{j,k}\bracket{\mathbf{z}_{i}}{\mathbf{z}_{j}} & \left(\left\{ \frac{d\mathbf{z}_{i}^{*}}{dt}\cdot\mathbf{z}_{j}-\frac{1}{2}\left[\mathbf{z}_{i}\cdot\frac{d\mathbf{z}_{i}^{*}}{dt}+\frac{d\mathbf{z}_{i}}{dt}\cdot\mathbf{z}_{i}^{*}\right]\right\} \bracket{\mathbf{a}_{i}}{\mathbf{a}_{j}}\right.\\
 & \quad\quad+\bracket{\dot{\mathbf{a}}_{i}}{\mathbf{a}_{j}}\\
 & \left.\quad\quad-\frac{\imath}{\hbar}\sum_{A,B}a_{i}^{*A}H_{AB}^{\text{ord}}\left(\mathbf{z}_{i}^{*},\mathbf{z}_{j}\right)a_{j}^{B}\right)\left(\Omega^{-1}\right)_{j,k}\bracket{\mathbf{z}_{k},\mathbf{a}_{k}}{\Psi}\label{eqn:time_dependence_zaPsi}
\end{split}
\end{equation}

Now one needs to fix the time-dependence for the trajectories that
guide the basis set. Each trajectory $i$ sits on an electronic state
$I_{i}$ and is propelled by the forces derived from the diagonal
element of the Hamiltonian, $H_{I_{i},I_{i}}^{\text{ord}}$:

\begin{equation}
\frac{d\mathbf{z}_{i}}{dt}=-\frac{\imath}{\hbar}\frac{\partial}{\partial\mathbf{z}^{*}}H_{I_{i},I_{i}}^{\text{ord}}(\mathbf{z}_{i}^{*},\mathbf{z}_{i})\label{eqn:time_dependence_z}
\end{equation}
These are just Newton's equations of motion (up to some additional
terms from reordering) when one combines position $\mathbf{q}$ and
momentum $\mathbf{p}$ into a single complex number $\mathbf{z}$.
They are integrated on the nuclear time scale (e.g. $\Delta t_{\text{nuc}}=0.1$)
fs.

The electronic coefficients follow 
\begin{equation}
\frac{da_{i}^{A}}{dt}=-\frac{\imath}{\hbar}\sum_{B}H_{AB}^{\text{ord}}(z_{i}^{*},z_{i})a_{i}^{B}\label{eqn:time_dependence_a}
\end{equation}
and are integrated on the electronic time scale (e. g. $\Delta t_{\text{elec}}=10^{-3}\Delta t_{\text{nuc}}$).
After each nuclear time step the trajectory can hop to a different
electronic state $J_{i}$ depending on the hopping probabilities that
are obtained from $\vec{a}_{i}(t)$ using Tully's original method
\cite{Tully_hopping_probabilities} or the improved modification \cite{Petric_hopping_probabilities}
of it, where the probabilities are calculated from the rates of change
of the quantum mechanical amplitudes: For the trajectory $i$ the
density matrix is computed as: 
\begin{equation}
\rho_{IJ}=a_{i}^{I}a_{i}^{J*}\quad\quad I,J\text{: electronic state labels}
\end{equation}
The probability to hop from state $I$ to state $J$ is calculated
from the diagonal elements and their derivatives \cite{Petric_hopping_probabilities}:
\begin{equation}
P_{I\to J}=\Theta(-\dot{\rho}_{II})\Theta(\dot{\rho}_{JJ})\frac{\left(-\dot{\rho}_{II}\right)\dot{\rho}_{JJ}}{\rho_{II}\sum_{K}\Theta(\dot{\rho}_{KK})\dot{\rho}_{KK}}\Delta t_{\text{nuc}}
\end{equation}
The formula can be rationalized as follows: A transition from $I$
to $J$ should only happen if the quantum population of $I$ decreases
and the quantum population on $J$ increases, $P_{I\to J}\propto\Theta(-\dot{\rho}_{II})\Theta(\dot{\rho}_{JJ})$,
it should be proportional to these changes, $P_{I\to J}\propto\left(-\dot{\rho}_{II}\right)\dot{\rho}_{JJ}$,
and it should go to zero as the time step decreases, $P_{I\to J}\propto\Delta t_{\text{nuc}}$.
The other terms ensure, that the conditional probability to hop to
any other state, given that the trajectory is on state $I$, is equal
to the change in probabilitiy over the time step $\Delta t_{\text{nuc}}$:
\begin{equation}
\rho_{II}\sum_{J}P_{IJ}=\Theta(-\dot{\rho}_{II})\left(-\dot{\rho}_{II}\right)\Delta t_{\text{nuc}}=\Theta(\Delta\rho_{II})\left(-\Delta\rho_{II}\right)
\end{equation}

Along each trajectory $i$ one also needs to integrate the classical
``action'' $S_{i}$ 
\begin{equation}
S_{i}=\int\frac{\imath\hbar}{2}\left(\mathbf{z}_{i}^{*}\cdot\frac{d\mathbf{z}_{i}}{dt}-\frac{d\mathbf{z}_{i}^{*}}{dt}\cdot\mathbf{z}_{i}\right)dt\label{eqn:time_dependence_S}
\end{equation}

Using the known time-dependence of $\mathbf{a}_{i}$, eqn. \ref{eqn:time_dependence_a},
the second line in eqn. \ref{eqn:time_dependence_zaPsi} can be replaced
by: 
\begin{equation}
\bracket{\dot{\mathbf{a}}_{i}}{\mathbf{a}_{j}}=\frac{\imath}{\hbar}\sum_{A,B}a_{i}^{A*}H_{AB}^{\text{ord}}(\mathbf{z}_{i}^{*},\mathbf{z}_{i})a_{j}^{B}
\end{equation}

The time derivative of the action can be used replace one derivative
in eqn. \ref{eqn:time_dependence_zaPsi}: 
\begin{equation}
-\frac{1}{2}\mathbf{z}_{i}^{*}\cdot\frac{d\mathbf{z}_{i}}{dt}=\frac{\imath}{\hbar}\frac{dS_{i}}{dt}-\frac{1}{2}\frac{d\mathbf{z}_{i}^{*}}{dt}\cdot\mathbf{z}_{i}
\end{equation}

Then, using eqns. \ref{eqn:time_dependence_z} and \ref{eqn:time_dependence_a},
one can rewrite eqn. \ref{eqn:time_dependence_zaPsi} into 
\begin{equation}
\begin{split}\frac{d}{dt}\bracket{\mathbf{z}_{i},\mathbf{a}_{i}}{\Psi} & =\frac{\imath}{\hbar}\frac{dS_{i}}{dt}\bracket{\mathbf{z}_{i},\mathbf{a}_{i}}{\Psi}+\frac{\imath}{\hbar}\sum_{j,k}\bracket{\mathbf{z}_{i}}{\mathbf{z}_{j}}\left(\bracket{\mathbf{a}_{i}}{\mathbf{a}_{j}}\frac{\partial}{\partial\mathbf{z}}H_{I_{i},I_{i}}^{\text{ord}}(\mathbf{z}_{i}^{*},\mathbf{z}_{i})(\mathbf{z}_{j}-\mathbf{z}_{i})\right.\\
 & \left.+\sum_{A,B}a_{i}^{A*}\left(H_{AB}^{\text{ord}}\left(\mathbf{z}_{i},\mathbf{z}_{i}\right)-H_{AB}^{\text{ord}}\left(\mathbf{z}_{i},\mathbf{z}_{j}\right)\right)a_{j}^{B}\right)\left(\Omega^{-1}\right)_{j,k}\bracket{\mathbf{z}_{k},\mathbf{a}_{k}}{\Psi}
\end{split}
\end{equation}

Now the coefficients $C_{i}(t)$ are introduced as 
\begin{equation}
\bracket{\mathbf{z}_{i},\mathbf{a}_{i}}{\Psi}=C_{i}(t)\exp\left(\frac{i}{\hbar}S_{i}(t)\right)
\end{equation}
with the time-dependence 
\begin{equation}
\frac{dC_{i}}{dt}e^{\frac{\imath}{\hbar}S_{i}}=\frac{d}{dt}\bracket{\mathbf{z}_{i},\mathbf{a}_{i}}{\Psi}-\frac{\imath}{\hbar}\frac{dS_{i}}{dt}\bracket{\mathbf{z}_{i},\mathbf{a}_{i}}{\Psi}
\end{equation}

The differential equation for these coefficients reads: 
\begin{equation}
\begin{split}\frac{dC_{i}}{dt}e^{\frac{\imath}{\hbar}S_{\mathbf{z}_{i}}}=-\frac{\imath}{\hbar}\sum_{j,k}\bracket{\mathbf{z}_{i}}{\mathbf{z}_{j}} & \left(\bracket{\mathbf{a}_{i}}{\mathbf{a}_{j}}\frac{\partial}{\partial\mathbf{z}}H_{I_{i},I_{i}}^{\text{ord}}(\mathbf{z}_{i}^{*},\mathbf{z}_{i})(\mathbf{z}_{i}-\mathbf{z}_{j})\right.\\
 & \left.+\sum_{A,B}a_{i}^{A*}\left(H_{AB}^{\text{ord}}(\mathbf{z}_{i}^{*},\mathbf{z}_{j})-H_{AB}^{\text{ord}}(\mathbf{z}_{i}^{*},\mathbf{z}_{i})\right)a_{j}^{B}\right)\left(\Omega^{-1}\right)_{j,k}C_{k}e^{\frac{\imath}{\hbar}S_{\mathbf{z}_{k}}}
\end{split}
\end{equation}
Since in this form the inverse of the overlap matrix is required,
a second set of coefficients $D_{j}(t)$ is introduced as: 
\begin{equation}
D_{j}(t)e^{\frac{\imath}{\hbar}S_{j}}=\sum_{k}\left(\Omega^{-1}\right)_{j,k}C_{k}e^{\frac{\imath}{\hbar}S_{k}}
\end{equation}
Which leads to: 
\begin{equation}
\begin{split}\frac{dC_{i}}{dt}e^{\frac{\imath}{\hbar}S_{i}}=-\frac{\imath}{\hbar}\sum_{j}\bracket{\mathbf{z}_{i}}{\mathbf{z}_{j}} & \left(\bracket{\mathbf{a}_{i}}{\mathbf{a}_{j}}\frac{\partial}{\partial\mathbf{z}}H_{I_{i},I_{i}}^{\text{ord}}(\mathbf{z}_{i}^{*},\mathbf{z}_{i})(\mathbf{z}_{i}-\mathbf{z}_{j})\right.\\
 & \left.+\sum_{A,B}a_{i}^{A*}\left(H_{AB}^{\text{ord}}(\mathbf{z}_{i}^{*},\mathbf{z}_{j})-H_{AB}^{\text{ord}}(\mathbf{z}_{i}^{*},\mathbf{z}_{i})\right)a_{\alpha}^{B}\right)D_{j}e^{\frac{\imath}{\hbar}S_{j}}
\end{split}
\end{equation}
The kernel of this differential equation is: 
\begin{equation}
\delta^{2}\mathcal{H}(\mathbf{z}_{i},\mathbf{a}_{i};\mathbf{z}_{j},\mathbf{a}_{j})=\bracket{\mathbf{z}_{i}}{\mathbf{z}_{j}}\left(\bracket{\mathbf{a}_{i}}{\mathbf{a}_{j}}\frac{\partial}{\partial\mathbf{z}}H_{I_{i},I_{i}}^{\text{ord}}(\mathbf{z}_{i}^{*},\mathbf{z}_{i})(\mathbf{z}_{i}-\mathbf{z}_{j})+\sum_{A,B}a_{i}^{A*}\left(H_{AB}^{\text{ord}}(\mathbf{z}_{i}^{*},\mathbf{z}_{j})-H_{AB}^{\text{ord}}(\mathbf{z}_{i}^{*},\mathbf{z}_{i})\right)a_{j}^{B}\right)\label{eqn:coupling_kernel}
\end{equation}

For each time step the coefficients $C_{i}$ are propagated according
to 
\begin{equation}
\frac{dC_{i}}{dt}e^{\frac{\imath}{\hbar}S_{i}}=-\frac{\imath}{\hbar}\sum_{j}\delta^{2}\mathcal{H}(\mathbf{z}_{i},\mathbf{a}_{i};\mathbf{z}_{j},\mathbf{a}_{j})D_{j}e^{\frac{\imath}{\hbar}S_{j}}
\end{equation}
and the guiding equations for $\mathbf{z}_{i}(t)$, $S_{i}(t)$ and
$\mathbf{a}_{i}(t)$ are propagated according to eqns. \ref{eqn:time_dependence_z},
\ref{eqn:time_dependence_S} (with a single step from $t$ to $t+\Delta t_{\text{nuc}}$)
and \ref{eqn:time_dependence_a} (from $t$ to $t+\Delta t_{\text{nuc}}$
with many smaller time steps of length $\Delta t_{\text{elec}}$).
During the integration of the electronic populations in eqn. \ref{eqn:time_dependence_a},
$H_{AB}^{\text{ord}}$ is interpolated linearly between $H_{AB}^{\text{ord}}(t)$
and $H_{AB}^{\text{ord}}(t+\Delta t_{\text{nuc}})$.

Then the next coefficients $D_{i}$ are determined by solving the
matrix equation 
\begin{equation}
\sum_{\beta}\Omega_{j,k}D_{k}e^{\frac{\imath}{\hbar}S_{k}}=C_{j}e^{\frac{\imath}{\hbar}S_{j}}\label{eqn:matrix_equation}
\end{equation}
In this scheme the inverse $\boldsymbol{\Omega}^{-1}$
is never calculated. Since coherent states are overcomplete, linear
dependencies between the moving basis vectors can lead to an almost
singular overlap matrix. For numerical stability eqn. \ref{eqn:matrix_equation}
is solved using the Lapack function ZHESVX\cite{lapack}. After
each time step trajectories may hop stochastically to another electronic
state with probability $P_{I\rightarrow J}$.

%The most expensive part is the integration of the electronic populations in eqn. \ref{eqn:time_dependence_a} and the solution of the matrix equation \ref{eqn:matrix_equation}.

Why does this propagation scheme work robustly? The quickly varying
degrees of freedom are absorbed into the guiding equations for the
basis functions, $\mathbf{z}_{i}(t),\mathbf{a}_{i}(t)$ and $S_{i}(t)$,
while the coupling between different basis functions, eqn. \ref{eqn:coupling_kernel},
always remains small \cite{CCS_method}: Coherent states are not orthogonal,
but their overlap decreases exponentially as they become more separated
in phase space, see eqn. \ref{eqn:zz_overlap}. Therefore first term
$\bracket{\mathbf{z}_{i}}{\mathbf{z}_{j}}$ in eqn. \ref{eqn:coupling_kernel}
keeps the coupling down for distant basis functions. For close basis
functions the coupling is also small, because of the second factor
in eqn. \ref{eqn:coupling_kernel}, that goes to zero for $\mathbf{z}_{i}\to\mathbf{z}_{j}$.

Some useful relations for calculating conserved quantities and quantum
probabilities are compiled in appendix \ref{sec:ccs_representation}.
More explicit formulae for the guiding equations can be found in appendix
\ref{sec:guiding_equations} and the inclusing of a time-dependent
electric field is discussed in appendix \ref{sec:field}.

% RESULTS

\section{Results}

\subsection{2D model for a Conical Intersection}

Ferretti et.al. \cite{ci_model} introduced a two-dimensional model
for a conical intersection (CI) in order to investigate to which extent
an ensemble of classical surface hopping trajectories can reproduce
the quantum mechanically exact solution. %Ferretti et.al. \cite{ci_model} examined the non-adiabatic dynamics at a two-dimensional model for a conical intersection (CI) and investigated to which extent an ensemble of classical trajectories can reproduce the quantum mechanically exact solution. Their trajectories were driven by the classical forces derived from the adiabatic potential energy surfaces and could jump stochastically between different adiabatic states accordint to Tull's surface hopping method. 

The model consists of two displaced 2-dimensional harmonic oscillators
that are coupled by a Gaussian off-diagonal element. The $2\times2$
diabatic potential matrix $\boldsymbol{V}(X,Y)$ has the form: 
\begin{eqnarray}
V_{11}(X,Y) & = & \frac{1}{2}K_{x}\left(X-X_{1}\right)^{2}+\frac{1}{2}K_{y}Y^{2}\\
V_{22}(X,Y) & = & \frac{1}{2}K_{x}\left(X-X_{2}\right)^{2}+\frac{1}{2}K_{y}Y^{2}+\Delta\\
V_{12}(X,Y) & = & V_{21}=\Gamma Y\exp\left(-\alpha(X-X_{3})^{2}-\beta Y^{2}\right)
\end{eqnarray}
The minima of the harmonic oscillators are located at $X_{1}=4.0$
and $X_{2}=3.0$, respectively. The coupling between the diabatic
states is strongest at $X_{3}=3.0$. The other constants are defined
as $K_{x}=0.02$, $K_{y}=0.1$, $\Delta=0.01$ and $\alpha=3.0$,
$\beta=1.5$. The masses belonging to the $X$ and $Y$ mode are set
to $M_{X}=20$ $000$ and $M_{Y}=6667$, respectively. The CI model
is investigated for different coupling strengths, for weak ($\Gamma=0.01$)
and strong coupling ($\Gamma=0.08$).

The initial wave packet is prepared as a Gaussian centered at $X_{0}=2.0$
and $Y_{0}=0.0$ on the first diabatic state, which on the left of
the conical intersection outside the interaction region coincides
with the second adiabatic state. Initially the diabatic wave function
is: 
\begin{eqnarray}
\xi_{1}(X,Y,t=0) & = & \frac{1}{\sqrt{\pi\Delta X\Delta Y}}\exp\left(-\frac{1}{2}\frac{(X-X_{0})^{2}}{\Delta X^{2}}-\frac{1}{2}\frac{Y^{2}}{\Delta Y^{2}}\right)\\
\xi_{2}(X,Y,t=0) & = & 0
\end{eqnarray}
with $\Delta X=0.150$ and $\Delta Y=0.197$.

%The wave packet slides down the energy surface until it reaches the conical intersection at $(X,Y) = (3.0,0.0)$ after 30 fs. 

Although the distribution of a large number of surface hopping trajectories
brings out the main aspects of the dynamics, some features defy a
semiclassical treatment: 
\begin{itemize}
\item In the ``shade'' of the conical intersection the probability density
is exactly zero. This fact cannot be explained semiclassically as
it originates from interference: If the nuclear wave packet moves
around a conical intersection the electronic wave function acquires
a Berry phase. The parts of the wave packet that flow around the left
and the right side of the conical intersection interfere destructively
because their phases are opposite. 
\item For large coupling strengths, the semiclassical treatment underestimates
the population transfer between the adiabatic states in comparison
with the exact quantum mechanical dynamics, which predicts that ``a
single crossing of a conical intersection is always a diabatic process''\cite{ci_model}. 
\item The comb-like interference pattern which develops behind the conical
intersection for strong coupling appears as a flat, broad plateau
without peaks or troughs in the semiclassical dynamics (see Fig. 3
for t=40 in \cite{ci_model}). 
\end{itemize}

\subsection{Numerical quantum dynamics}

For comparison the time-dependent Schrödinger equation was solved
on an equidistant two-dimensional grid using the second order differences
(SOD) method: 
\begin{equation}
\Psi(\mathbf{x},t+\Delta t)\approx\Psi(\mathbf{x},t-\Delta t)-2\frac{\imath}{\hbar}\hat{H}\Psi(\mathbf{x},t)\Delta t\quad\quad\quad\text{ with }\hat{H}=V(\mathbf{x})+T(\hat{\mathbf{p}})
\end{equation}
Since the potential energy operator $V(\mathbf{x})$ is diagonal in
the position representation and the kinetic energy operator $T(\hat{\mathbf{p}})$
is diagonal in the momentum representation, the action of $V$ on
the wave function was computed in real space and the action of $T$
in momentum space: 
\begin{equation}
\hat{H}\Psi(t)=V(\mathbf{x})\Psi(\mathbf{x},t)+\mathcal{F}^{-1}\left\{ T(\hat{\mathbf{p}})\mathcal{F}\left\{ \Psi(\mathbf{x},t)\right\} \right\} 
\end{equation}
The Fast Fourier transform allows to switch quickly between the two
representations in each propagation step\cite{fft_qm}, $\tilde{\Psi}(\mathbf{p},t)=\mathcal{F}\left\{ \Psi(\mathbf{x},t)\right\} $.
The grid covered the range $-1\le X\le8$ and $-3\le Y\le3$ with
150 points in both $X$ and $Y$ direction and a time step of $\Delta t=0.01$
fs was used to propagate the wave function for 100 fs.

\subsection{Dynamics with surface hopping coupled coherent states}

The width parameter for the coherent states was set to $\gamma=25.0$
so that the size of the coherent states resemble the spatial extension
of the initial wave packet.

\paragraph{weak coupling ($\Gamma=0.01$):}

\begin{figure}[h!]
\includegraphics[width=0.8\textwidth]{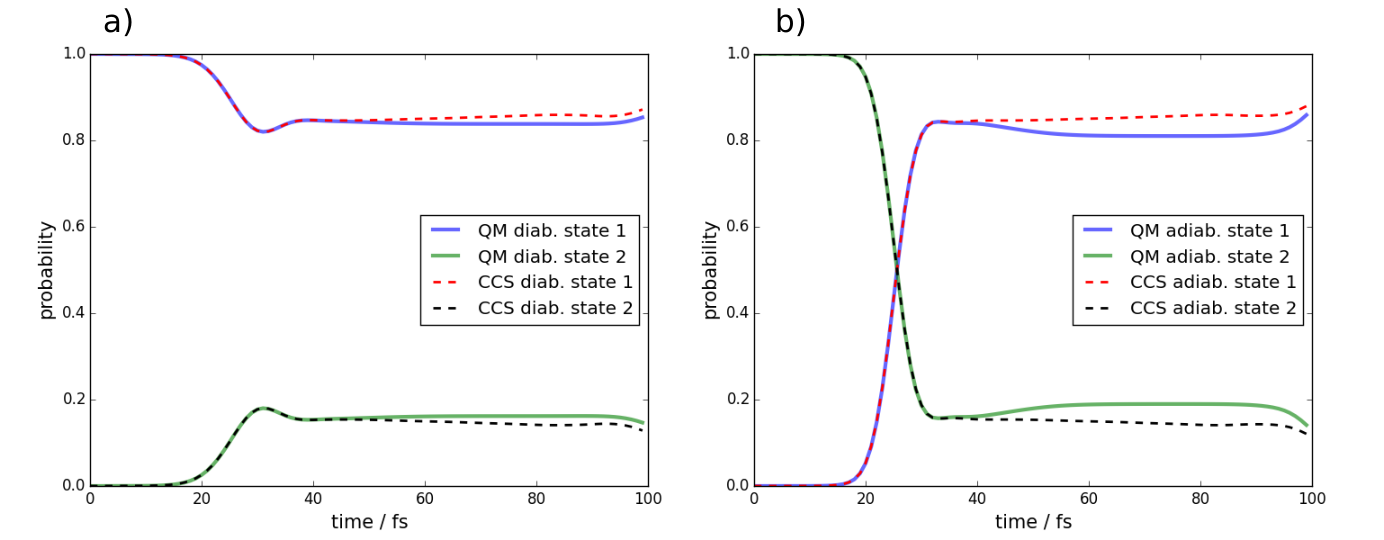} \caption{\textbf{State probabilities} for weak ($\Gamma=0.01$) coupling in
the \textbf{a)} diabatic and \textbf{b)} adiabatic picture. Numerically
exact (solid), CCS dynamics (dashed).}
\label{fig:ci-weak-populations} 
\end{figure}

\begin{figure}[h!]
\includegraphics[width=1\textwidth]{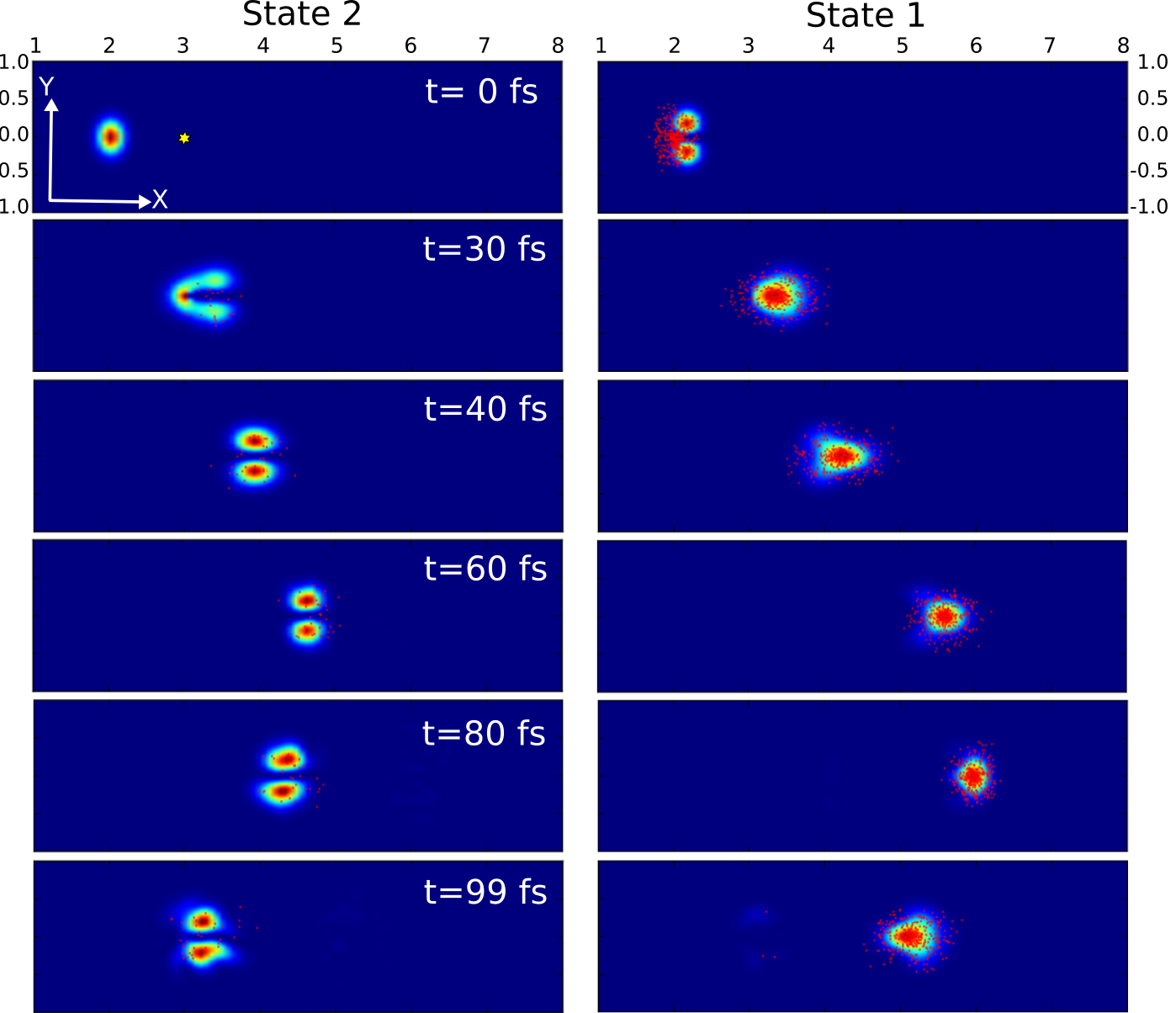}
\caption{\textbf{Adiabatic wave functions} $\vert\Psi_{1,2}(X,Y,t)\vert^{2}$
for weak ($\gamma=0.01$) coupling at different times. The position
of the CI is marked by a small yellow star in the first frame. The
red dots indicate the centers of the coherent states, which guide
the moving basis set. The wavepackets were transformed from the diabatic
representation (in which the simulation is performed) to the adiabatic
representation, while trajectories are shown for diabatic states.
Therefore at time $t=0.0$ fs all red dots are located on the diabatic
state 1, whereas the wavepacket starts out in the adiabatic state
2. For better contrast, the color range extends from $\vert\Psi_{1,2}\vert_{\text{min}}^{2}$
(blue) to $\vert\Psi_{1,2}\vert_{\text{max}}^{2}$ (red) in each image;
at $t=0$ fs, $\vert\Psi_{2}(0)\vert_{\text{max}}^{2}=3.6$, $\vert\Psi_{1}(0)\vert_{\text{max}}^{2}=8\cdot10^{-5}$
and at $t=99$ fs, $\vert\Psi_{2}(99)\vert_{\text{max}}^{2}=0.2$,
$\vert\Psi_{1}(99)\vert_{\text{max}}^{2}=1.6$.}
\label{fig:ci-weak-frames} 
\end{figure}

Initial conditions for $N_{\text{traj}}=250$ trajectories were sampled
from the Wigner distribution. The equations of motion for the trajectories
and the equations for the coupling between them were integrated with
a time step of $\Delta t_{\text{nuc}}=0.01$ fs. In each nuclear time
step the equation for the electronic amplitudes were integrated with
a time step of $\Delta t_{\text{elec}}=5\cdot10^{-6}$ fs. Initially
151 trajectories participate in the computation of the coupling coefficients.
This number is enlarged to 250 as the trajectories disperse on the
potential energy surface. Coupling all trajectories from the start,
when they are still very closely packed, could lead to a singular
overlap matrix. Towards the end of the simulation the CCS results
deviate a little bit from the exact ones, but this can be amended
by increasing the number of trajectories further. Fig. \ref{fig:ci-weak-populations}
depicts total state probabilities in the adiabatic and the diabatic
picture. Snapshots of the wavepackets at different times are shown
in Fig.\ref{fig:ci-weak-frames}.

\FloatBarrier

\paragraph{strong coupling ($\Gamma=0.08$):}

To reproduce the numerically exact results, much more trajectories
are needed for the strong coupling regime than for the weak one.

\begin{figure}[h!]
\includegraphics[width=0.8\textwidth]{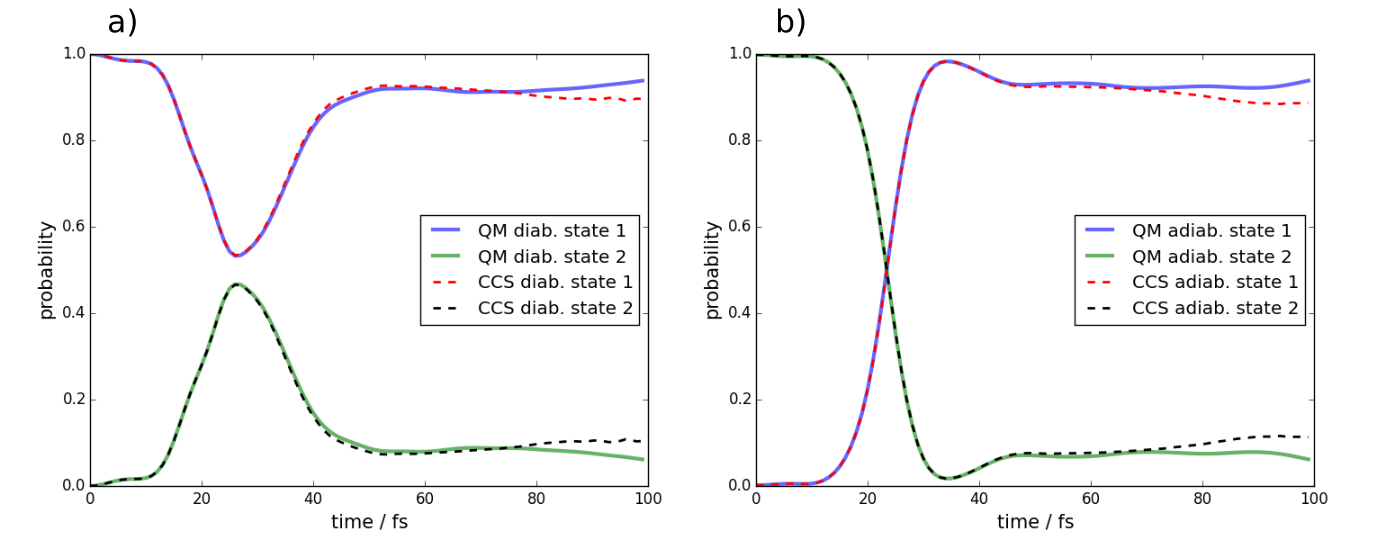} \caption{\textbf{State probabilities} for strong ($\Gamma=0.08$) coupling
in the \textbf{a)} diabatic and \textbf{b)} adiabatic picture. Numerically
exact (solid), CCS dynamics (dashed).}
\label{fig:ci-strong-populations} 
\end{figure}

\begin{figure}[h!]
\includegraphics[width=1\textwidth]{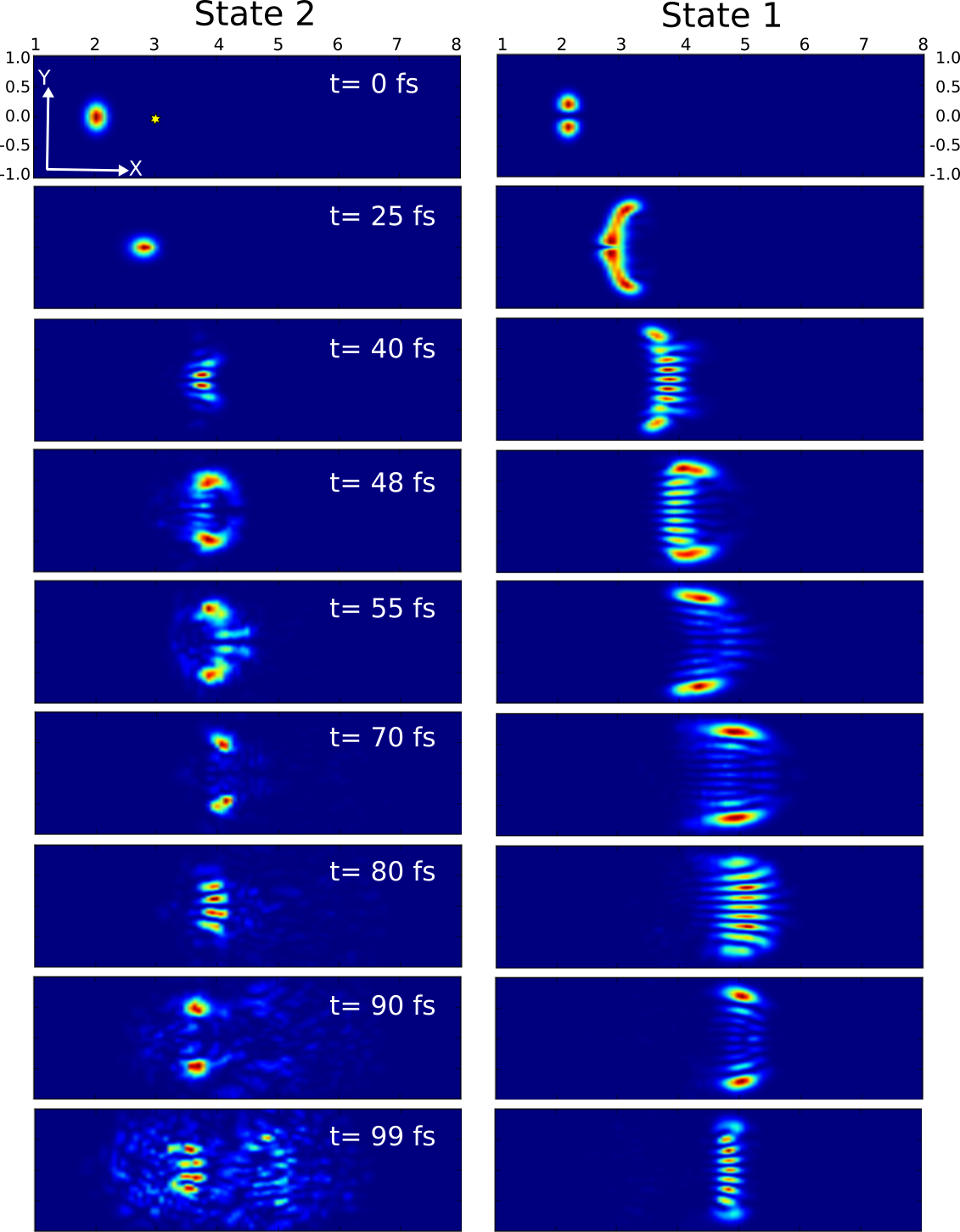}
\caption{\textbf{Adiabatic wave functions} $\vert\Psi_{1,2}(X,Y,t)\vert^{2}$
for strong ($\gamma=0.08$) coupling at different times.}
\label{fig:ci-strong-frames} 
\end{figure}

Initial conditions for 1500 trajectories are sampled from $W(q,p)^{1/3}$.
Sampling from the cubic root of the Wigner distribution makes the
initial trajectory distribution more diffuse, so that the trajectories
do not overlap too much. A nuclear time step of $\Delta t_{\text{nuc}}=0.01$
fs and an electronic time step of $\Delta t_{\text{elec}}=3\cdot10^{-6}$
was used. The resulting total state probabilities are shown in Fig.\ref{fig:ci-strong-populations},
snapshots of the wavepacket evolution are shown in Fig.\ref{fig:ci-strong-frames}.

Interestingly, most of the time is spent in integrating the electronic
populations for the surface hopping procedure, so the cost of CCS
dynamics is not so different from usual surface hopping. The limitation
is that for CCS dynamics the potential energy surface has to be known
globally (e.g. in the form of a force field) while for surface hopping
local knowledge of the energy, gradient and non-adiabatic couplings
is enough.

\FloatBarrier

\paragraph{Trajectory Populations:}

\begin{figure}[h!]
\includegraphics[width=0.8\textwidth]{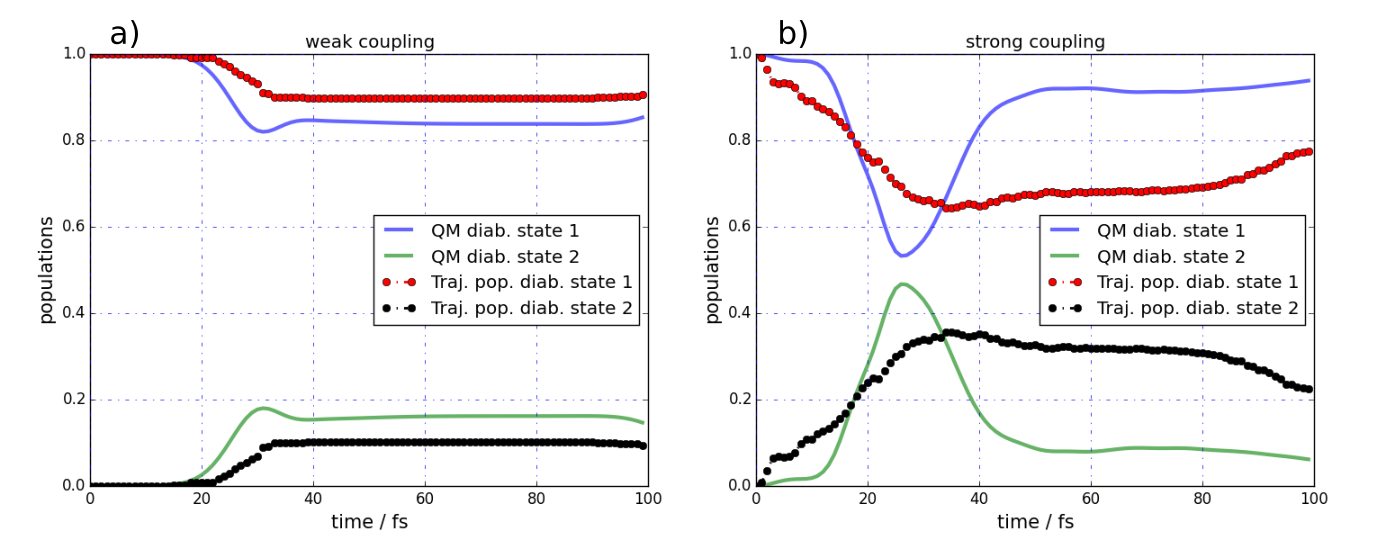}
\caption{\textbf{Comparison between trajectory populations and QM probabilities}
for \textbf{a)} weak and \textbf{b)} strong coupling. The trajectory
populations (dots) are obtained by counting the number of trajectories
on each diabatic state. In \textbf{b)} the trajectory populations
deviate from the quantum populations, because the initial distribution
were sampled not from the Wigner distribution $W(q,p)$ but from $W(q,p)^{1/3}$.}
\label{fig:ci-qmpop-pop} 
\end{figure}

It is also instructive to look at the populations of the guiding trajectories
on the two diabatic states (see Fig.\ref{fig:ci-qmpop-pop}). In the
case of weak coupling the trajectory populations underestimate the
transfer of population between the diabatic states. In the case of
strong coupling they look completely different: The initial conditions
were sampled from the cubic root of the Wigner function, and therefore
represent a different semiclassical wavepacket. Using a different
initial distribution is a valid trick, since the trajectories only
function as a basis, which can be distributed at will as long as it
covers the region where the wavepacket passes through. The quantum
populations still agree very well for both coupling strengths (see
Figs. \ref{fig:ci-weak-populations} and \ref{fig:ci-strong-populations}).

\section{Conclusions and Outlook}

By solving the Schrödinger equation in the basis of surface hopping
coherent states the complex interference effects around a conical
intersection can be fully reproduced. This is not surprising as no
approximations have been made apart from using a finite basis set.
Therefore, the method could serve as an alternative to numerically
exact grid-based propagation schemes in more than 3 dimensions, provided
the diabatic potentials can be expressed in a form, for which the
matrix elements between coherent states can be computed analytically.
This is a severe limitation that does not affect direct dynamics schemes,
where only adiabatic gradients and non-adiabatic couplings are required.
On the other hand, although methods for direct quantum dynamics are
sometimes claimed to be exact, a convergence to the exact result is
not guaranteed, if matrix elements are approximated for compatibility
with electronic structure calculations.

Further work will focus on developing building blocks for analytical
molecular potentials. Molecular diabatic potentials can be expanded
into terms depending only on bond lengths, bond angles, dihedrals
etc.; conical intersections or avoided crossing can be modelled by
Gaussians placed on the off-diagonals. The averaging integrals would
have to be worked out for a set of force field-like terms from which
potential energy surfaces for larger molecules can be constructed
in the spirit of empirical valence bond theory\cite{warshel_evb,ms_evb}.
This would allow to perform numerically exact quantum dynamics on
model potentials to investigate the photochemistry of small molecules.

%The pattern depends strongly on the correct accumulation of the dynamic and geometric phase along each CCS trajectory. %It is known, that a sign change occurs when the electronic wavefunction is transported adiabatically around a conical intersection.

%- Total probabilities - integrated quantities - are much easier%- Intereference %- emphasize importance of exact matrix elements, globally known potential energy surfaces and couplings.%- potential energy surfaces are quadratic but coupling is not%- direct dynamics where couplings between gaussians are approximated by quantities (energies, gradients, non-adiabatic coupling vectors) available locally, should in general not lead to the correct intereference pattern. Instead the presence of the CI %Methods that use only local quantities such as gradients, hessians, non-adiabatic coupling vectors%- 

\section{Acknowledgements}

We thank Stewart Reed for making his CCS code publicly
available, which has proven helpful in debugging the code developed
for this work. The financial support by the European Research Council
(ERC) Consolidator Grant ``DYNAMO'' Grant Nr. is gratefully acknowledged.

\appendix
%dummy comment inserted by tex2lyx to ensure that this paragraph is not empty

\section{Appendix}

\subsection{Wavefunction in the CCS representation}

\label{sec:ccs_representation} The basis of surface hopping coherent
states offers a very compact representation for a multi-dimensional
wavefunction that can be delocalized over many electronic states.
For convenience a few useful relations are list here: 
\begin{itemize}
\item The wave function is the following superposition of the basis functions:
\begin{equation}
\ket{\Psi}=\sum_{i}\ket{\textbf{z}_{i},\mathbf{a}_{i}}D_{i}e^{\frac{\imath}{\hbar}S_{i}}\label{eqn:Psi_coherent_states}
\end{equation}

\item Its norm is given by: 
\begin{equation}
\bracket{\Psi}{\Psi}=\sum_{i}C_{i}^{*}D_{i}
\end{equation}

\item The total energy, which in the absence of an external field should
be a conserved quantity, is 
\begin{equation}
E_{\text{tot}}=\bra{\Psi}\hat{H}\ket{\Psi}=\sum_{i}\left(D_{i}e^{\frac{\imath}{\hbar}S_{i}}\right)^{*}\sum_{j}\bracket{\mathbf{z}_{i}}{\mathbf{z}_{j}}\left(\sum_{A,B}a_{i}^{A*}H_{AB}^{\text{ord}}\left(\mathbf{z}_{i}^{*},\mathbf{z}_{j}\right)a_{j}^{B}\right)\left(D_{j}e^{\frac{\imath}{\hbar}S_{j}}\right),\label{eqn:etot}
\end{equation}

\item and the quantum probability to be on state $I$ can be obtained as:
\begin{equation}
p_{I}=\vert\bracket{I}{\Psi}\vert^{2}=\sum_{i,j}\left(D_{i}a_{i}^{I}e^{\frac{\imath}{\hbar}S_{i}}\right)^{*}\bracket{\mathbf{z}_{i}}{\mathbf{z}_{j}}\left(D_{j}a_{j}^{I}e^{\frac{\imath}{\hbar}S_{j}}\right)
\end{equation}

\end{itemize}

\subsection{Guiding equations for a diabatic hamiltonian}

\label{sec:guiding_equations} For a diabatic hamiltonian with the
form 
\begin{equation}
\hat{H}_{AB}=\delta_{AB}T(\hat{\mathbf{p}})+V_{AB}(\mathbf{x})=\delta_{AB}\sum_{d=1}^{N_{\text{dim}}}\frac{\hat{p}_{d}^{2}}{2m_{d}}+V_{AB}(x_{1},\ldots,x_{d})
\end{equation}
the kinetic energy can be reordered algebraically to give 
\begin{equation}
T^{\text{ord}}(\mathbf{z}_{i}^{*},\mathbf{z}_{j})=-\frac{\gamma\hbar^{2}}{4}\sum_{d=1}^{N_{\text{dim}}}\frac{1}{m_{d}}\left[\left(z_{j,d}-z_{i,d}^{*}\right)^{2}-1\right].
\end{equation}
with the gradient 
\begin{equation}
\frac{\partial}{\partial z_{d}^{*}}T^{\text{ord}}(\mathbf{z},\mathbf{z})=\frac{\imath\hbar^{2}\gamma}{m_{d}}\Im(z_{d})
\end{equation}

The equations of motion for the action, the complex position vector
and the electronic amplitudes of a trajectory become 
\begin{eqnarray}
\frac{dS}{dt} & = & =\sum_{d=1}^{N_{\text{dim}}}\frac{\hbar^{2}\gamma}{m_{d}}\left(\Im(z_{d})\right)^{2}+\Re(\mathbf{z})\cdot\frac{\partial V_{II}^{\text{ord}}}{\partial\mathbf{z}}\\
\frac{dz_{d}}{dt} & = & \frac{\hbar\gamma}{m_{d}}\Im(z_{d})-\frac{\imath}{\hbar}\frac{\partial V_{II}^{\text{ord}}}{\partial z_{d}}\\
\frac{da^{A}}{dt} & = & -\frac{\imath}{\hbar}\sum_{B}\left[T^{\text{ord}}(\mathbf{z}^{*},\mathbf{z})\delta_{AB}+V_{AB}^{\text{ord}}(\mathbf{z}^{*},\mathbf{z})\right]a^{B}
\end{eqnarray}
where $\Re$ and $\Im$ denote the real and imaginary part and $I$
is the current electronic state of the trajectory.

\subsection{Interaction with light}

\label{sec:field} 
\begin{figure}[h!]
\includegraphics[width=0.6\textwidth]{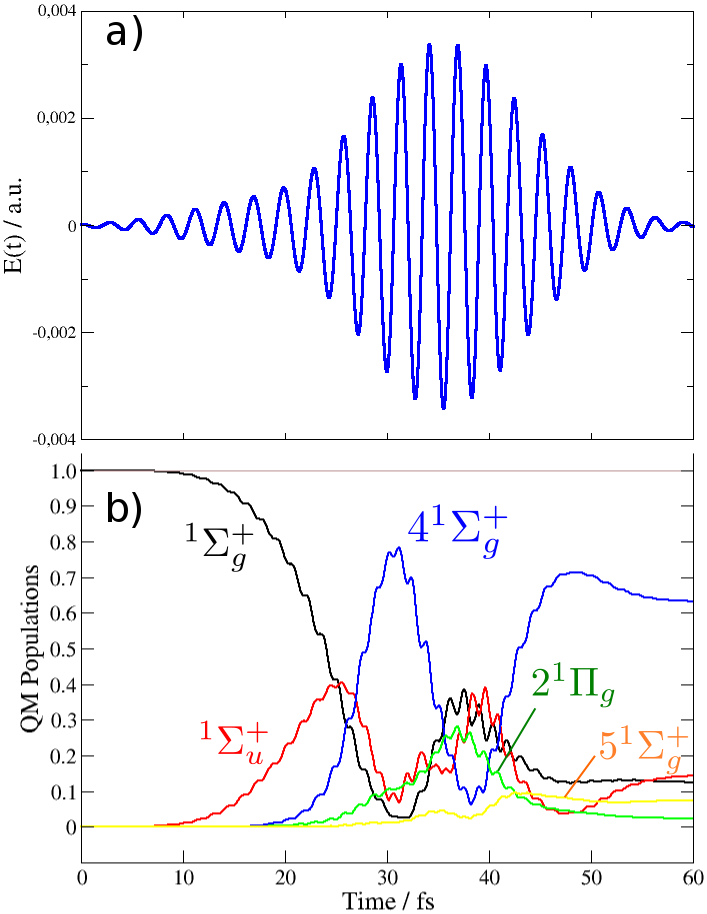} \caption{\textbf{Field induced dynamics in K$_{2}$.} Analytical functions
of the bond length were fitted to the diabatic potential energy surfaces
and transition dipoles of K$_{2}$ at the CASSCF/MRCI level\cite{Petric_hopping_probabilities}.
The wavepacket, initially prepared on the ground state, was propagated
using 300 coherent state trajectories in the presence of a short light-pulse.
Although the problem is effectively one-dimensional, the trajectories
were propagated including all 6 cartesian coordinates. \textbf{a)}
Time-dependence $E(t)$ of the electric pulse (same as in Fig 2a of
Ref. \cite{Petric_hopping_probabilities}), \textbf{b)} Quantum populations
obtained with surface hopping coherent states.}
\label{fig:k2_populations_field} 
\end{figure}

Interaction with a time-dependent external electric
field can be included to simulate pump-probe experiments or coherent
control. For simplicity, the vectorial nature of the electric field
is neglected, and the dot product between the field vector and the
transition dipole, $\vec{E}\cdot\vec{\mu}$, is replaced by $E\mu$.
The additional time-dependent part of the Hamiltonian reads: 
\begin{equation}
\hat{H}_{AB}^{\text{field}}(t)=E(t)V_{AB}^{\text{field}}(\mathbf{x})
\end{equation}
where $E(t)$ only depends on time and $V_{AB}^{\text{field}}(\mathbf{x})$
represents the magnitudes of the transition dipoles between the electronic
states $A$ and $B$ (which depend on the nuclear geometries). Since
the time-dependence is limited to $E(t)$, the integrals for ``reordering''
$V_{AB}^{\text{field}}$ have to be calculated only once every nuclear
time-step and remain constant during the integration of the electronic
amplitudes. The hopping probabilities of the coherent states are driven
by the electric field as in the field-induced surface hopping (FISH)
method \cite{Petric_hopping_probabilities}. In Fig. \ref{fig:k2_populations_field}
the quantum-populations of the electronic states in the $K_{2}$ molecule
during excitation with a shaped pulse are shown.

%%%%%%%%%%%%%%% BIBLIOGRAPHY %%%%%%%%%%%%%%%%%%%%%%%\providecommand{\refin}[1]{\\ \textbf{Referenced in:} #1} 

% COUPLED COHERENT STATES

%%%%%%%%% citations in the introduction %%%%%%%%%%%%%

% classical surface hopping MD


\begin{thebibliography}{10}
\bibitem{marx_md_on_excited_states_review} Doltsinis,~N.;\ \ Marx,~D.;
First principles molecular dynamics involving excited states and non-adiabatic
transitions. \textit{J. Theor. Comput. Chem.,} \textbf{2002,} \textsl{1,}
319-349.


% Hydrogen Tunneling

\bibitem{voth_H_tunneling} Lobaugh,~J.;\ \ Voth,~G.; The quantum
dynamics of an excess proton in water. \textit{J. Chem. Phys.,} \textbf{1996,}
\textsl{104,} 2056-2069.

\bibitem{molecular_Akharonov_Bohm} Mead,~A.;\ \ Truhlar,~D.;
On the determination of Born-Oppenheimer nuclear motion wave functions
including complications due to conical intersections and identical
nuclei. \textit{J. Chem. Phys.,} \textbf{1979,} \textsl{70,} 5.


% Conical Intersections

\bibitem{ci_book} \textsl{Conical Intersections: Electronic Structure,
Dynamics and Spectroscopy,} edited by Domcke,~W.;\ \ Yarkony,~D.;\ \ Köppel,~H.;
\textit{World Scientific Publishing, Singapore,} \textbf{2004}.

\bibitem{diabolic_ci} Yarkony,~D.; Diabolic conical intersections.
\textit{Rev. Mod. Phys.,} \textbf{1996,} \textsl{68,} 4.

\bibitem{geometric_phase_phenol} Abe,~M.;\ \ Ohtsuki,~Y.;\ \ Fujimura,~Y.;\ \ Lan,~Zh.;\ \ Domcke,~W.;
Geometric phase effects in the coherent control of the branching ratio
of photodissociation products of phenol. \textit{J. Chem. Phys.,}
\textbf{2006,} \textsl{124,} 224316.

\bibitem{CCS_method} Shalashilin,~D.;\ \ Child,~M.; The phase
CCS approach to quantum and semiclassical molecular dynamics for high-dimensional
systems. \textit{Chem. Phys.} \textbf{2004,} \textsl{304,} 103-120.

\bibitem{frozen_gaussians} Heller,~E.; Frozen Gaussians: A very
simple semiclassical approximation. \textit{J. Chem. Phys.,} \textbf{1981,}
\textsl{75,} 2923.

\bibitem{CCS_ehrenfest} Shalashilin,~D.; Nonadiabatic dynamics with
the help of multiconfigurational Ehrenfest method: Improved theory
and fully quantum 24D simulation of pyrazine. \textit{J. Chem. Phys.}
\textbf{2010,} \textsl{132,} 244111.

\bibitem{Tully_hopping_probabilities} Tully,~J.; Molecular dynamics
with electronic transitions. \textit{J. Chem. Phys.} \textbf{1990,}
\textsl{93,} 1061.

\bibitem{surface_hopping_gaussians} Horenko,~I.;\ \ Salzmann,~Ch.;\ \ Schmidt,~B.;\ \ Schütte,Ch.;
Quantum-classical Liouville approach to molecular dynamics: Surface
hopping Gaussian phase-space packets. \textit{J. Chem. Phys.,} \textbf{2002,}
\textsl{117,} 11075.

\bibitem{mctdh_book} Meyer,H-D.;\ \ Gatti,~F.;\ \ Worth,G.;
Multdimensional Quantum Dynamics - MCTDH Theory and Applications.
\textit{Wiley-VCH,} \textbf{2009}.

\bibitem{mctdh_zundel_cation} Vendrell,~O.;\ \ Gatti,~F.;\ \ Meyer,~H-D.;
Dynamics and Infrared Spectroscopy of the Protonated Water Dimer.
\textit{Angew. Chem. Int. Ed.,} \textbf{2007,} \textsl{46,} 6918-6921.

\bibitem{ab_initio_multiple_spawning} Ben-Nun,~M.;\ \ Mart\'{i}nez,~T.;
Ab initio quantum molecular dynamics. \textit{Adv. Chem. Phys.} \textbf{2002,}
\textsl{121,} 439.

\bibitem{aims_ci} Mart\'{i}nez,~T.; Ab initio molecular dynamics
around a conical intersection: Li(p)+H$_{2}$. \textit{Chem. Phys.
Lett.,} \textbf{1997,} \textsl{272,} 139-147.

\bibitem{aims_tunneling} Ben-Nun,~M.;\ \ Mart\'{i}nez,~T.; A
multiple spawning approach to tunneling dynamics. \textit{J. Chem.
Phys.} \textbf{2000,} \textsl{112,} 6113.

\bibitem{ab_initio_multiple_cloning} Makhov,~D.;\ \ Glover,~W.;\ \ Mart\'{i}nez,~T.;\ \ Shalashilin,~D.;
Ab initio multiple cloning algorithm for quantum non-adiabatic molecular
dynamics. \textit{J. Chem. Phys.} \textbf{2014,} \textsl{141,} 054110.

\bibitem{Na3F_qmdynamics} Heitz,~M.;\ \ Durand,~G.;\ \ Spiegelman,~F.;
Ultrafast excited state dynamics of the Na$_{3}$F cluster: Quantum
wave packet and classical trajectory calculations compared to experimental
results. \textit{J. Chem. Phys.,} \textbf{2004,} \textsl{121,} 9906-9916.

\bibitem{wavepacket_Jacobi} Dixon,~R.; A Three-dimensional Time-dependent
Wavepacket Calculation for Bound and Quasi-bound Levels of the Ground
State of HCO: Resonance Energies, Level Widths and CO Product State
Distributions. \textit{J. Chem. Soc. Faraday Trans.,} \textbf{1992,}
\textsl{88,} 2575-2586.

\bibitem{CCS_cartesian} Reed,~S.;\ \ González Mart\'{i}nez,~L.;\ \ Rubayo
Soneira,~J.;\ \ Shalashilin,~D.; Cartesian coupled coherent states
simulations: Ne n Br2 dissociation as a test case. \textit{J. Chem.
Phys.} \textbf{2011,} \textsl{134,} 054110.

\bibitem{ccs_sampling} Shalashilin,~D.;\ \ Child,~M.; Basis set
sampling in the method of coupled coherent states: Coherent state
swarms, trains and pancakes. \textit{J. Chem. Phys.,} \textbf{2007,}
\textsl{128,} 054102.

\bibitem{aims_molpro} Levine,~B.;\ \ Coe,~J.;\ \ Virshup,~A.;\ \ Mart\'{i}nez,~T.;
Implementation of ab initio multiple spawning in the MOLPRO quantum
chemistry package. \textit{J. Chem. Phys.,} \textbf{2008,} \textsl{347,}
3-16.

\bibitem{mctdh_nonadiabatic} Worth,~G.;\ \ Meyer,~H.;\ \ Köppel,~H.;\ \ Cederbaum,~L.;\ \ Burghardt,~I.;
Using the MCTDH wavepacket propagation method to describe multimode
non-adiabatic dynamics. \textit{Int. Rev. in Phys. Chem.,} \textbf{2008,}
\textsl{27,} 569-606.

\bibitem{burghardt_vMCG} Worth,~G.;\ \ Robb,~M.;\ \ Burghardt,~I.;
A novel algorithm for non-adiabatic direct dynamics using variational
Gaussian wavepackets. \textit{Faraday Discuss.,} \textbf{2004,} \textsl{127,}
307-323.

\bibitem{truhlar_fit_NH3} Nangia,~Sh.;\ \ Truhlar,~D. Direct
calculation of coupled diabatic potential-energy surfaces for ammonia
and mapping of a four-dimensional conical intersection seam. \textit{J.
Chem. Phys.,} \textbf{2006,} \textsl{124,} 124309.

\bibitem{klauder_book} Klauder,~J.;\ \ Skagerstam,~B.; Coherent
States - Applications in Physics and Mathematical Physics. \textit{World
Scientific,} \textbf{1985.}

\bibitem{lapack} Anderson,~E.;\ \ Bai,Z.;\ \ Bischof,~C.;\ \ Blackford,~S.;\ \ Demmel,~J.;\ \ Dongarra,~J.;\ \ Du
Croz,~J.;\ \ Greenbaum,~A.;\ \ Hammarling,~S.;\ \ McKenney,~A.;\ \ Sorensen,~D.
LAPACK Users' Guide. \textit{Society for Industrial and Applied Mathematics,}
\textbf{1999,} \textsl{3rd edition.}

\bibitem{ci_model} Ferretti,~A.;\ \ Granucci,~G.;\ \ Lami,~A.;\ \ Persico,~M.;\ \ Villani,~G.
Quantum mechanical and semiclassical dynamics at a conical intersection.
\textit{J. Chem. Phys.,} \textbf{1996,} \textsl{104,} 5517.

\bibitem{footnote_moving_basis} To appreciate intuitively why moving
basis sets can be very efficient imagine a TV crew is to cover the
Tour de France. If they would like to cover the entire sports event
with stationary cameras, they would have to place a camera every few
hundred meters along the entire route. Most of the camera would not
see anything interesting for most of the time. If they mount cameras
on cars and follow the peloton, relatively few cameras are needed
depending on how closely the cyclists stay together.

\bibitem{Petric_hopping_probabilities} Petersen,~J.;\ \ Mitri\'{c},~R.;
Electronic coherence within the semiclassical field-induced surface
hopping method: strong field quantum control in $K_{2}$. \textit{Phys.
Chem. Chem. Phys.} \textbf{2012,} \textsl{14,} 8299-8306.

\bibitem{fft_qm} Kosloff,~D.;\ \ Kosloff,~R. A Fourier Method
Solution for the Time Dependent Schrödinger Equation as a Tool in
Molecular Dynamics. \textit{J. Comput. Phys.,} \textbf{1983,} \textsl{52,}
35-53.

\bibitem{warshel_evb} Warshel,~A.;\ \ Weiss,~R.; An Empirical
Valence Bond Approach for Comparing Reactions in Solutions and in
Enzymes. \textit{J. Am. Chem. Soc.,} \textbf{1980,} \textsl{102,}
6218-6226.

\bibitem{ms_evb} Schmitt,~U.;\ \ Voth,~G.; Multistate Empirical
Valence Bond Model for Proton Transport in Water. \textit{J. Phys.
Chem. B,} \textbf{1998,} \textsl{102,} 29.\end{thebibliography}
\end{document}